\documentstyle[prl,aps]{revtex}
\newcommand{\BEQ}{\begin{equation}}
\newcommand{\EEQ}{\end{equation}}
\newcommand{\BEA}{\begin{eqnarray}}
\newcommand{\EEA}{\end{eqnarray}}

\renewcommand{\d}{{\rm d }}
\renewcommand{\b}{{\hat \beta }}
\newcommand{\T}{{\hat T }}

\newcommand{\nn}{\nonumber \\}
\begin{document}
\draft
\title{The Marriage Problem and the Fate of Bachelors}
\author{ Th.~M.~Nieuwenhuizen}
\address{Van der Waals-Zeeman Instituut, University of Amsterdam
\\ Valckenierstraat 65, 1018 XE Amsterdam, The Netherlands}
\date{October 8, 1997; \today}
\maketitle
\begin{abstract}
In the marriage problem, a variant of the bi-parted matching problem,
each member has a `wish-list' expressing 
his/her preference for all possible partners; this list consists
of random, positive real numbers drawn from a certain distribution. 
One searches the lowest  cost for the society, at the risk of 
breaking up pairs in the course of time. 
Minimization of a global cost function (Hamiltonian) is performed 
with statistical mechanics techniques at a finite fictitious
temperature. 

The problem is generalized to include bachelors, needed in particular
when the groups have different size, and polygamy. 
Exact solutions are found for the optimal solution ($T=0$).
The entropy is found to vanish quadratically in $T$. 
Also other evidence is found that the replica symmetric solution is exact,
implying at most a polynomial degeneracy
of the optimal  solution.
 
Whether bachelors occur or not, depends not only on their
intrinsic qualities, or lack thereof, 
but also on global aspects of the chance for 
pair formation in society.
\end{abstract}

\section{Introduction}
The stable marriage is a well-known optimization problem
``women'', and pairs are formed between them.
Many other applications exist, such as the job-market 
(employers and employees), the housing market, etc.
Each man ranks the women according to his preference, number one
being his most preferred candidate; likewise,
each woman ranks the men. The cost function for a configuration
of marriages is the sum of the rankings of the actual partners
on the two lists.
In the past most studies have been concerned with algorithms for finding 
 stable pairing solutions. Indeed, given the rules according to
which marriages may be ended and started at new, a stable
situation may be reached where it brings no  benefit to break up
existing relations for starting other ones.
For a recent study of this subject, see Om\'ero et al ~\cite{Omero}.

Since there is intrinsically quenched 
disorder (the lists, see below), many such stable solutions are actually
meta-stable in the sense that their cost function is not minimal.
If the rules for breaking pairs are then softened, the society will
go to a state of lower cost, i.e., more happiness.
In some sense this is just what happens in reality. Indeed, due to
changes in morality, marriage rules are weakened in the course of time. 
One century ago it was in most societies both morally and economically
almost impossible to separate married couples, 
whereas nowadays it is a generally accepted phenomenon
in most Western and several Eastern societies. 
In line of the principle of freedom of the individual,
many American movies and television soap series nowadays even 
advocate that a person has the right to improve his/her own situation,
no matter what
cost this brings to the present partner, to the surroundings of the
aimed new partner, or to society.
To give a proper description
of the total happiness of society, one should then, however, also 
take into account the cost for breaking up existing families.
Such a more complete description would, naturally, put limits to
individual freedom.

Let us consider a group $A$ of $N$ men and a group $B$ of $M$ woman.
In reality the ratio $\nu=M/N$ will slightly exceed unity.
The man $i$ has a wish-list, that is to say
he evaluates all the women in group $B$ and ranks them into a preference
order. He will prefer, of course, the number one on his list, but that woman
may not have reciprocal intentions. 
Our task is to find a best global arrangement
in which the total ``happiness'' is optimal. This calls for a cost function
which we shall refer to as the Hamiltonian. In each arrangement preferably
one man is paired to one woman. However, the possibility of remaining single 
is needed if the groups have a different number of members. Also polygamy 
can be  permitted with assigned penalty weights.

Globally stable solutions can be only found for small systems using
exact enumeration methods. 
It is not hard to imagine that, as the system becomes larger, many 
conflicting wishes make a satisfactory solution harder to find.
A  systematical analysis is called for, which is conveniently 
done in terms of a partition sum at a fictitious temperature $T$;
finally the $T\to 0$ limit should be taken.
There is a connection with spin glasses where many 
different but roughly equivalent states occur. It was this connection 
which led in the mid-eighties to the replica analysis of several 
combinatorial problems~\cite{MPV}, such as the bi-parted matching problem by
Orland~\cite{Orland}, 
the mono-parted matching problem and the traveling salesman problem
by M\'ezard and Parisi~\cite{MezardParisi}.

The wish list can be implicitly given for the man $i$. 
For a given distribution he draws 
a (random) set of real positive numbers  $\ell_{ij}^a$, that gives the
energy cost for pairing with  woman $j$. 
This set can in principle be arranged into a increasing order, with the most 
preferred partner having the least cost.
Likewise, each woman $j$ has
a different (random) list $\ell_{ij}^b$ giving the energy cost 
to form a pair with man $i$. 

In standard studies the cost of a configuration of marriages is the
sum of rankings on the lists. For matching two groups of $N$ members
each person therefore brings an integer cost $C$, $1\le C\le N$, that
equals the rank of his actual partner on his wish list. 
These integer costs are not so realistic, however. Number ten on a
large list of candidates will almost be just as acceptable as number one, 
and not ten times less. For this reason we will follow
Orland and M\'ezard-Parisi, who
have considered non-integer costs, $\ell_{ij}$.

The total cost for pair $(i,j)$ is then
$\ell_{ij}=\alpha \ell_{ij}^a+(1-\alpha)\ell_{ij}^b$. Note that in general 
$\ell_{ij}^a\neq \ell_{ij}^b$ since intentions are not in general reciprocal. 
$\alpha$ characterizes
another asymmetry of the problem: when $\alpha=1$ the cost function only
represents men's interest, with that of women totally ignored, vice versa
$\alpha=0$ women's wish prevails. Besides these extremal men or women 
dominated situations, the case $\alpha=1/2$ represents the 
symmetrical (ideal) balance between the two groups. 
In society $\alpha$ is slightly tilt to men's advantage.
   
Apart from this, if the man $i$ remains single, this costs an energy $A_i$.
Basically, if $A_i>\min_j \ell_{ij}^a$ then it is profitable for him
to form a pair. Likewise, the cost for woman $j$ to remain single
is $B_j$. We shall also allow for polygamy in a grand-canonical fashion. 
If man $i$ marries $k\ge 2$ women, there is an additional 
energy cost $(k-1)\mu_{1i}$, where $\mu_{1i}$ plays the role of 
a chemical potential. Similarly there is a cost $(k-1)\mu_{2j}$ when woman
$j$ marries $k\ge 2$ man. Also the $\mu$'s can be taken as random
variables, or may just be fixed.

The goal of the problem is to find the best possible solution, i.e., 
the one that has lowest energy (``cost'') for the whole society. 
It remains to be seen in how far this is the best solution
for a typical individual, and which dynamics 
has the best state as stable solution.

For equally large groups $(M=N)$ and and no singles $(A_i=B=\infty)$ 
nor polygamy $(\mu_{1i}=\mu_{2j}=\infty)$
this marriage problem is closely related to the bipartite matching problem
 studied by Orland ~\cite{Orland} and by M\'ezard and 
Parisi ~\cite{MezardParisi}. The major difference lies in
the asymmetry of energy costs $\ell_{ij}^a\neq\ell_{ij}^b$, 
the new features such like being single or polygamy presents 
also new questions.  Our analysis will follow
these approaches as much as it is possible.
    
The setup of the paper is as follows. In section II we describe some 
simple scaling aspects of the problem. In section III we describe a
detailed replica analysis. 
In section IV we make a low $T$ analysis. We close with a discussion
and summary. In the appendix we consider the energy fluctuations.

\section{Scaling analysis of the problem}

A state is fully characterized by the set of numbers $\{n_{ij}\}$,
where $n_{ij}=1$ if the pair $(i,j)$ is formed and zero else. 
The total number of partners of man $i$ is $n_i=\sum_j n_{ij}$. Its
typical value is $n_i=1$ (man $i$ finds a woman/a woman finds him); 
however, $n_i=0$ (man $i$ remains single) or $n_i\ge 2$ (he is a polygamist)
may also occur.
In a similar way $m_j\equiv \sum_i n_{ij}$ describes the fate of woman $j$.
The interesting case is where $n_i$ and $m_j$ are annealed variables,
so that both possibilities can be weighted thermally.

The energy of a state is

\BEQ
{\cal H}=\sum_{ij}n_{ij}\ell_{ij}+\sum_i A_i\delta_{n_i,0}
+\sum_i\sum_{k=2}^\infty (k-1)\mu_{1i}\delta_{n_i,k}
+\sum_j B_j\delta_{m_j,0}
+\sum_j\sum_{k=2}^\infty (k-1)\mu_{2j}\delta_{m_j,k}
\EEQ
Let us first consider the scaling for the pairs formed. The cost for pair
$(i,j)$ is
\BEQ
\ell_{ij}=\alpha\ell_{ij}^{(a)}+(1-\alpha)\ell_{ij}^{(b)}
\EEQ
In order to form a pair, this combined value should be as low as possible.
First consider $\alpha=1$. Then typically the lowest $\ell_{ij}^{(a)}$
is $\ell_{min}^{(a)}$ is given by 
\BEQ
\int_0^{\ell_{min}^{(a)}}
p_a(\ell)d\ell\approx \frac{1}{N}
\EEQ
Since we shall take 
\BEQ p_a(\ell)=\frac{\ell^r_a e^{-\ell}}{\Gamma(r_a+1)}\EEQ
 this
implies $\ell_{min}^{(a)}=[\Gamma(r_a+1)/N]^{1/(r_a+1)}$.
The pairing energy would be equal
\BEQ
U_{pairs}=N\ell_{min}^{(a)}\sim N^{r_a/(r_a+1)}
\EEQ
Typically, a man would find a woman that is high on the top of his list.
The women would accept this, since their wishes have no effect what so ever.

Now, if woman's preference is also taken into account ($\alpha<1$) things
change. For $\alpha=1/2$ the distribution for the total $\ell$  is:
\BEQ
p(\ell)=\frac{2^{r+1}\ell^re^{-2\ell}}{\Gamma(r+1)}\qquad r=r_a+r_b+1
\EEQ
while for general $\alpha$ the small-$\ell$ behavior is
\BEQ \label{pdeell}
p(\ell)\approx \frac{\ell^r }{\alpha^{r_a+1}(1-\alpha)^{r_b+1}\Gamma(r+1)}
 \qquad (\ell\to 0)
\EEQ
The typical value $\ell_{typ}\sim \ell_{min}
\sim N^{-1/(r+1)}$ now implies that the typical partner is located
on position $k$ of the wish list related as 
$\int_0^{\ell_{typ}} d\ell p_a(\ell) =k/N$, yielding
$k\sim N^{(r_b+1)/(r+1)}$. For the symmetric case $r_a=r_b=(r-1)/2$ 
this becomes $k\sim\sqrt{N}$. 
So for the two-side weighted case the wishes of each
individual are quite moderately fulfilled: 
the ideal partner is not within reach.
She/he has typically has different wishes, which excludes formation of the
pair. For the bulk of pairs the partners are only
moderately biased towards each other. This is the price society has to
pay for individual freedom.

The ground state energy will scale as $U_0\sim N\ell_{typ}\sim
N^{r/(r+1)}$. This implies that bachelors and polygamist s only play 
an interesting role when their costs 
($A_i, B_j,\mu_{1,i}$ and  $\mu_{2,j}$)
scale as $ U_0/N\sim N^{-1/(r+1)}$. 
If the costs have a larger magnitude, 
these options do not occur provided the groups have equal size; 
if they are smaller, no pairs will be formed.
Notice that these scalings are not universal to the individual,
but depend on the exponent $r$ entering the distribution
(\ref{pdeell}) of costs for pairings in the whole societies.

\section{The partition sum and replica's}

We consider the problem in a statistical mechanics approach. We thus 
consider the sum over all configurations of Boltzmann factors $\exp(-\beta E)$
at given temperature $T=1/\beta$.
The optimal solution has lowest energy; it will be dominant in the limit
$T\to 0$. The statistical approach also yields the  entropy, related to the
degeneracy of the optimal solution, as well as the average numbers of
married man etc.

To simplify notation we first take sure values for $A_i$, $B_j$, $\mu_{1\,i}$,
$\mu_{2\,j}$.
The partition sum can be written as
\BEA
Z&=&\prod_{i=1}^N\prod_{j=1}^M\sum_{n_{ij}=0}^1\sum_{n_i,m_j=0}^\infty  
     e^{-\beta H}\nonumber\\
 &=&\sum_{n_{ij}=0}^1 \sum_{n_i,m_j=0}^\infty 
e^{-\beta\sum_{i=1}^N \{ (A+\mu_1)\delta_{n_i,0}+(n_i-1)\mu_1\}}
e^{-\beta\sum_{j=1}^M\{ (B+\mu_2)\delta_{m_j,0}+(m_j-1)\mu_2\}}
\nonumber\\
&e&^{-\beta \sum_{ij}n_{ij}\ell_{ij}}
\prod_{i=1}^N\delta(n_i-\sum_{j=1}^Mn_{ij})
\prod_{j=1}^M\delta(m_j-\sum_{i=1}^Nn_{ij})
\EEA
where the $\delta$'s are Kronecker $\delta$-functions.
Repeatedly using  the integral representation
\BEQ
\delta(a-b)=\int_0^{2\pi} \frac{\d\lambda}{2\pi} e^{i\lambda(a-b)}
\EEQ
 one arrives 
at a form where the sums can be carried out. After doing that one gets
\BEQ
Z=\prod_{i=1}^N\int_0^{2\pi} \frac{d\lambda_i}{2\pi}
\frac{e^{-\beta A}+(1-e^{-\beta (A+\mu_1)})e^{i\lambda_i}}
{1-e^{-\beta\mu_1+i\lambda_i}}  
  \prod_{j=1}^M\int_0^{2\pi} \frac{d\mu_j}{2\pi}
\frac{e^{-\beta B}+(1-e^{-\beta (B+\mu_2)})e^{i\mu_j}}
{1-e^{-\beta\mu_2+i\mu_i}}  
\prod_{i,j=1}^{N,M} (1+e^{-i\lambda_i-i\mu_j-\beta\ell_{ij}})
\EEQ
In order to calculate the quenched average free energy one 
employs the replica trick:
\BEQ \overline{F}=-T\overline{\ln Z}=-T\lim_{n\to 0} \frac{\overline{Z^n}-1}{n}
\EEQ
One thus replicates the partition sum $n$ times and uses for each 
pair  $(i,j)$
\BEQ
\prod_{\alpha=1}^n(1+e^{-i\lambda_i^\alpha-i\mu_j^\alpha-\beta\ell_{ij}})
=1+\sum_{p=1}^n \sum_{1\le\alpha_1< \cdots<\alpha_p\le n}
e^{-i(\lambda_i^{\alpha_1}+\cdots+\lambda_i^{\alpha_p})
-i(\mu_j^{\alpha_1}+\cdots+\mu_j^{\alpha_p})-p\beta\ell_{ij}}
\EEQ
Now the quenched average over $\ell_{ij}$ can be carried out. One gets
\BEQ
 \overline{e^{-p\beta \ell_{ij}}}=
\int d\ell_{ij}^ap(\ell_{ij}^a)d\ell_{ij}^bp(\ell_{ij}^b)
e^{-p\beta (\alpha\ell_{ij}^a+(1-\alpha)\ell_{ij}^b)}\equiv\frac{g_p}{N}
\EEQ
This defines
\BEQ \label{gp=}
g_p=\frac{N}{(1+\alpha p\beta)^{r_a+1}(1+(1-\alpha) p\beta)^{r_b+1}}
\approx \frac{N}{(p\beta) ^{1+r}\alpha^{r_a+1}(1-\alpha)^{r_b+1}}
\EEQ
where $r=r_a+r_b+1$. In terms of $r$ our $g_p$ has a similar form 
as in the matching problem~\cite{Orland}\cite{MezardParisi}.

We shall confine ourselves to low temperatures, where we scale temperature 
with $N$ such that
\BEQ\label{That=}
\hat T=T\left(\frac{N}{\alpha^{r_a+1}(1-\alpha)^{r_b+1}}\right)^{1/(r+1)}
\EEQ
is of order unity. This implies that
\BEQ
g_p\approx \left(\frac{\hat T}{p}\right)^{r+1}
\EEQ 
is also of order unity. We then get to leading order in $N$ 
\BEQ
\overline{ Z^n}=
\prod_{i,\alpha}\int \frac{d\lambda_i^\alpha}{2\pi}
\frac{e^{-\beta A}+(1-e^{-\beta (A+\mu_1)})e^{i\lambda_i ^\alpha}}
{1-e^{-\beta\mu_1+i\lambda_i^\alpha}}
\prod_{j,\alpha}\int \frac{d\mu_j^\alpha}{2\pi}
\frac{e^{-\beta B}+(1-e^{-\beta (B+\mu_2)})e^{i\mu_j^\alpha}}
{1-e^{-\beta\mu_2+i\mu_j^\alpha}}  
\,\, e^S\EEQ
with
\BEQ
S=\frac{1}{N}{\sum_{ij}\sum_{p=1}^n g_p
\quad\sum_{1\le\alpha_1< \cdots<\alpha_p\le n}
e^{-i(\lambda_i^{\alpha_1}+\cdots+\lambda_i^{\alpha_p})
-i(\mu_j^{\alpha_1}+\cdots+\mu_j^{\alpha_p})}}
\EEQ
The $i,j$-dependence of $\exp S$ can be decoupled by a 
Hubbard-Stratonovich-type transformation
\BEA
e^S&=&\prod_{p=1}^n\prod_{\{\alpha_r\}}
\int \d P_{\alpha_1\cdots\alpha_p}\d Q_{\alpha_1\cdots\alpha_p}
\exp\left[-N\sum_{p=1}^n\frac{1}{g_p}
\sum_{1\le\alpha_1< \cdots<\alpha_p\le n}
Q_{\alpha_1\cdots\alpha_p}P_{\alpha_1\cdots\alpha_p}\right]\nonumber\\
&\times&\exp\sum_{p=1}^n\sum_{1\le\alpha_1< \cdots<\alpha_p\le n}
\{P_{\alpha_1 \cdots \alpha_p}\sum_{i=1}^N
e^{-i(\lambda_i^{\alpha_1}+\cdots+\lambda_i^{\alpha_p})}
+Q_{\alpha_1\cdots\alpha_p}\sum_{j=1}^M
e^{-i(\mu_j^{\alpha_1}+\cdots+\mu_j^{\alpha_p})}\}
\EEA
Now all $i,j$ indices are decoupled. The $\lambda_i$ integrals 
 all yield the same factor. The replicated partition sum thus becomes equal
to a $P,Q$ integral of a function $\exp[-N\Phi(P,Q)]$.
Because $N$ is large, we may approximate the integral by its saddle point 
value. Therefore the replicated free energy $F_n=-T\ln\overline{Z^n}$ 
is just equal to this saddle point value. We thus obtain
\BEQ\label{bFn}
\frac{\beta F_n}{N}=\sum_{p=1}^n\frac{1}{g_p}
\sum_{1\le\alpha_1< \cdots<\alpha_p\le n}
Q_{\alpha_1\cdots\alpha_p}P_{\alpha_1\cdots\alpha_p}
- z^a_n-\nu z^b_n\EEQ
where the latter two objects are defined as 
\BEQ e^{z^a_n}= 
\prod_{\alpha=1}^n\left[\int_0^{2\pi}\frac{d\lambda_\alpha}{2\pi}
\frac{e^{-\beta A}+(1-e^{-\beta (A+\mu_1)})e^{i\lambda_\alpha}}
{1-e^{-\beta\mu_1+i\lambda_\alpha}}  \right] 
\exp\sum_{p=1}^n\sum_{1\le\alpha_1< \cdots<\alpha_p\le n}
P_{\alpha_1\cdots\alpha_p}
e^{-i(\lambda_{\alpha_1}+\cdots+\lambda_{\alpha_p})} \EEQ
and similarly 
\BEQ e^{z^b_n}=  
 \prod_{\alpha=1}^n\left[\int_0^{2\pi}\frac{d\mu_\alpha}{2\pi}
\frac{e^{-\beta B}+(1-e^{-\beta (B+\mu_2)})e^{i\mu_\alpha}}
{1-e^{-\beta\mu_2+i\mu_\alpha}}  
\right]
\exp\sum_{p=1}^n\sum_{1\le\alpha_1< \cdots<\alpha_p\le n}
Q_{\alpha_1\cdots\alpha_p}
e^{-i(\mu_{\alpha_1}+\cdots+\mu_{\alpha_p})} \EEQ

Following Orland and M\'ezard-Parisi we assume that the relevant
saddle point has replica symmetry, viz. 
$Q_{\alpha_1\cdots\alpha_p}=Q_p$, 
$P_{\alpha_1\cdots\alpha_p}=P_p$.
Then the $\lambda_\alpha$ integrals can be replaced by contour integrals over
$z_\alpha=\exp(-i\lambda_\alpha)$, and the integral can be evaluated
from the poles at $z_\alpha=0$ and $\exp(-\beta\mu_1)$. 
This yields
 \BEA\label{ezan=} e^{z^a_n}&=&e^{n\beta\mu_1}
\prod_{\alpha=1}^n\sum_{z_\alpha=0,e^{-\beta\mu_1}}
\left[(-1+e^{-\beta A-\beta\mu_1})\delta_{z_\alpha,0}+
\delta_{z_\alpha,e^{-\beta\mu_1}}\right]
\exp\sum_{p=1}^nP_p\sum_{1\le\alpha_1<\cdots<\alpha_p\le n}
z_{\alpha_1}\cdots z_{\alpha_p}\nonumber\\
&=&e^{n\beta\mu_1}
\sum_{k=0}^n\left(n\atop k\right)
(-1+e^{-\beta A-\beta \mu_1})^k
\exp\sum_{p=1}^{n-k}P_p\left(n-k\atop p\right)e^{-\beta \mu_1 p}
\nonumber\\
&=&e^{n\beta\mu_1}
\sum_{k=0}^\infty\left(n\atop k\right)(-1)^k
(1-e^{-\beta A-\beta \mu_1})^k
\exp\sum_{p=1}^\infty (-1)^pP_p\frac{\Gamma(k+p-n-i0)}{\Gamma(k-n-i0)p!}
e^{-\beta \mu_1 p}
\EEA
where $i0=i\epsilon$ with $\epsilon\to 0$ first.
In the last step extension of the sums to $\infty$ was allowed 
because the extra terms are identically zero.
So far $n$ has been integer. Now we can continue this expression
for $n\to 0$. There are two types of contributions: the $k=0$
term equals $1+{\cal O}(n)$, while the $k\ge 1$ are all 
${\cal O}(n)$.
Indeed, for those terms one uses
\BEQ\label{noverk}
 \left(n\atop k\right)=\frac{\Gamma(n+i0+1)}{\Gamma(n-k+1)k!}
=\frac{\Gamma(n+1)\sin(\pi(k-n-i0))\Gamma(k-n-i0)}{\pi k!}\to
\frac{n(-1)^{k-1}}{k}(1-n\psi(k)+n\psi(1)+\cdots)
\EEQ
In the limit $n\to 0$ one thus gets
\BEQ\label{zan=} {z^a_n}=nz_a+n^2z_a^{(2)}+{\cal O}(n^3)\EEQ 
The result for $z_a$ is presented in eq. 
(\ref{z_a}), while  $z_a^{(2)}$ is presented in eq. (\ref{za2=})

Likewise $z_b\equiv\lim_{n\to 0}{z^b_n}/n$ is given in eq. (\ref{z_b}).

Further we find, using eq. (\ref{noverk}),
\BEA\label{QPn}
\sum_{p=1}^n\frac{1}{g_p}
\sum_{1\le\alpha_1< \cdots<\alpha_p\le n}
Q_{\alpha_1\cdots\alpha_p}P_{\alpha_1\cdots\alpha_p}&=&
\sum_{p=1}^\infty\frac{Q_pP_p}{g_p}
\left({n\atop p}\right)
\to
n\sum_{p=1}^\infty\frac{(-1)^{p-1}}{p}\frac{Q_pP_p}{g_p}
\EEA

\subsection{The quenched free energy}

Now the limit $n\to 0$ can be taken simply. 
The quenched average 
free energy  $F=\lim_{n\to 0}F_n/n$ thus follows from 
eq. (\ref{bFn}) as
\BEQ \label{bFN}
\frac{\beta F}{N}=\sum_{p=1}^\infty\frac{Q_pP_p}{g_p}
\frac{(-1)^{p-1}}{p}-
z_a(A,\mu_1)- \nu z_b(B,\mu_2)
\EEQ
with
\BEQ\label{z_a}
z_a(A,\mu_1)=
\beta\mu_1+\sum_{p=1}^\infty\frac{(-1)^{p-1}}{p}P_pe^{-p\beta\mu_1}
-\sum_{k=1}^\infty\frac{(1-e^{-\beta (A+\mu_1)})^{k}}{k}
\exp\left\{\sum_{p=1}^\infty
P_p\frac{(-1)^{p}\Gamma(k+p)}{\Gamma(k)p!}
e^{-p\beta\mu_1}\right\}
\EEQ
and
\BEQ\label{z_b}
z_b(B,\mu_2)=\beta\mu_2+\sum_{p=1}^\infty
\frac{(-1)^{p-1}}{p}Q_pe^{-p\beta\mu_2}
-\sum_{k=1}^\infty\frac{(1-e^{-\beta (B+\mu_2)})^{k}}{k}
\exp\left\{\sum_{p=1}^\infty(-1)^{p}Q_p\frac{\Gamma(k+p)}{\Gamma(k)p!}
e^{-p\beta\mu_2}\right\}
\EEQ
In section 2 we have discussed the possibility that
$A_i$, $B_j$, $\mu_{1\,i}$ and $\mu_{2\,j}$ are independent random
variables. For that case the analysis goes along the same steps; one only has
to average the present expressions for $z_a$ and $z_b$ with respect to 
these variables.

In  the case where these variables are non-random, and only the 
pair costs $\ell_{ij}^{(a,b)}$ are random, 
the saddle point equations read:

\BEQ\label{Qpgen=}
Q_p=g_pe^{-p\beta\mu_1}+pg_pe^{-p\beta\mu_1}\sum_{k=1}^\infty
\frac{[1-e^{-\beta(A+\mu_1)}]^k}{k}  \frac{\Gamma(k+p)}{\Gamma(k)p!}
e^{-P(k)}
\EEQ
and
\BEQ\label{Ppgen=}
P_p=\nu g_pe^{-p\beta\mu_2}+\nu pg_pe^{-p\beta\mu_2}\sum_{k=1}^\infty
\frac{[1-e^{-\beta(B+\mu_2)}]^k}{k}  \frac{\Gamma(k+p)}{\Gamma(k)p!}
e^{-Q(k)}
\EEQ
where $-P(k)$ and $-Q(k)$ are the arguments of the exponent
 in the $k$'th term of $z_a$ and $z_b$, respectively. 

We can compare this with the Orland and M\'ezard-Parisi results.
In the limit $\mu_{1,2} \to \infty$ only large-$k$ terms are relevant.
Introducing $\xi=ke^{-\beta\mu_{1,2}}$ we can go to an integral representation.
Adding and subtracting to $z_a$ the expression $\sum_{k=1}^\infty \exp(-\xi)/k$
we regularize the small $\xi$ behavior. We find that the $\mu$ contributions
 cancel, and obtain in the limit
\BEA\label{xi-int}
\frac{\beta F}{N}&=&\sum_{p=1}^\infty\frac{Q_pP_p}{g_p}\frac{(-1)^{p-1}}{p}-
\int_0^\infty \frac{d\xi}{\xi} 
[e^{-\xi}-\exp\sum_{p=1}^\infty \frac{P_p(-\xi)^p}{p!}
e^{-\xi e^{-\beta A}}]\nonumber\\
&-&\nu  \int_0^\infty \frac{d\xi}{\xi} 
[e^{-\xi}-\exp\sum_{p=1}^\infty \frac{Q_p(-\xi)^p}{p!}
e^{-\xi e^{-\beta B}}]
\EEA
In absence of bachelors ($A,B\to \infty$) and the limit
of equal group sizes ($\nu=M/N\to 1$) this equation reduces
to the result of   Orland and Mezard-Parisi.
Equation (\ref{xi-int}) was also derived by us using their 
method of taking derivatives rather than summing over residues.

The saddle point equations follow from eqs. (\ref{bFN}), (\ref{z_a}),
and (\ref{z_b}).
\BEA\label{Qeqnxi}
Q_p=\frac{pg_p}{p!}
\int_0^\infty d\xi e^{-\xi e^{-\beta A}}\xi^{p-1}
\exp\sum_{p=1}^\infty \frac{P_p(-\xi)^p}{p!}
\EEA
and
\BEA\label{Peqnxi}
P_p=\nu\frac{pg_p}{p!} \int_0^\infty d\xi
e^{-\xi e^{-\beta B}}\xi^{p-1}
\exp\sum_{p=1}^\infty \frac{Q_p(-\xi)^p}{p!}
\EEA

\section{Low temperatures}

From eq. (\ref{That=}) one sees that $g_p=(p\hat \beta)^{-(1+r)}$
 is of order unity if we
scale temperatures with $N$ such that $\hat T\sim  T N^{1/(1+r)}$
remains fixed.
As discussed in section II, the free energy is not 
extensive but scales as $F=-T\log Z=-N^{-1/(r+1)}\hat T
\log Z=N^{r/(r+1)}\hat F$ with intensive $\hat F$.

For having non-trivial behavior 
 one also needs that $A$ scales properly with
$N$, $A=aN^{-1/(r+1)}$ so that $\beta A=\hat \beta a$,
and similarly $\mu_{1,2}=\hat\mu_{1,2}N^{-1/(r+1)}$.

Inserting $\xi =\exp(\hat\beta \ell)$
 we can now introduce the functions $\hat P(\ell)=P(\xi)$, etc. 
To simplify notation, we shall
however, again denote $\hat P(\ell)$ by $P(\ell)$, etc. Thus 
\BEQ 
 P(\ell)=\sum_{p=1}^\infty \frac{(-1)^{p-1}P_p}{p!}e^{p\hat\beta \ell}
\qquad 
 Q(\ell)=\sum_{p=1}^\infty \frac{(-1)^{p-1}Q_p}{p!}e^{p\hat\beta \ell}
\EEQ
At finite $\hat T$ they satisfy
\BEQ\label{PQell}
 P(\ell)=\nu  \int_{-\infty}^\infty \d y 
 B(\ell+y)e^{- Q(y)}e^{-e^{\hat\beta(y-a)}}
\EEQ
which for $\hat T\to 0$ reduces to
\BEQ\label{PT=0}
P(\ell)=\nu\int_{-\ell}^a \d y B(\ell+y)e^{-Q(y)} 
\EEQ
Similarly
\BEQ\label{QT=0}
 Q(\ell)= 
 \int_{-\infty}^\infty \d y 
 B(\ell+y)e^{- P(y)}e^{-e^{\hat\beta(y-b)}}
\to
\int_{-\ell}^b \d y B(\ell+y)e^{-P(y)} 
\EEQ
with
\BEA\label{Br=}
 B(\ell)&=&\hat T^r
\sum_{p=1}^\infty \frac{(-1)^{p-1}e^{p\hat\beta\ell}}{p^rp!p!}
=\frac{\hat\beta}{\Gamma(r+1)}\int_0^\infty \d y y^r 
\sum_{p=1}^\infty \frac{(-1)^{p-1}e^{p\hat\beta(\ell-y)}}{p!(p-1)!}
\nonumber\\&=&
\frac{\hat\beta}{\Gamma(r+1)}\int_0^\infty \d y y^r 
e^{\hat\beta(\ell-y)/2}J_1(2e^{\hat\beta(\ell-y)/2})
=\frac{1}{\Gamma(r+1)}\int_{-\infty}^{\hat\beta\ell} d\eta (\ell-\hat T\eta)^r 
e^{\eta/2}J_1(2e^{\eta/2}) \nonumber\\
&\to& \left\{ 
{\frac{\ell^r}{\Gamma(r+1)}\qquad \ell>0} \atop
     {0\qquad\qquad \ell<0}
\right.\EEA
Here $J_1$ is a Bessel function.

\subsection{Observables}
The internal energy scales with $N$ as
\BEQ U=\frac{T}{\hat T}\hat U= \frac{1}{N}
\left({\alpha^{r_a+1}(1-\alpha)^{r_b+1}}\right)^{1/(r+1)}\hat U
\EEQ
The scaled internal energy 
$\hat U=\partial (\hat\beta\hat F)/\partial\hat\beta$
consists of three terms
\BEQ
\hat U=\hat U_{ab}+\hat U_a+\hat U_b
\EEQ
The pair energy is
\BEA
\frac{1}{N}\hat U_{ab}&=&\frac{1+r}{\hat \beta}
\sum_{p=1}^\infty\frac{Q_pP_p}{g_p}\frac{(-1)^{p-1}}{p}\nonumber\\
&=&(1+r) 
\int_{-\infty}^\infty d\ell\hat P(\ell)
e^{-e^{\beta(\ell-a)}}e^{-\hat P(\ell)}\to
(1+r)\int_{-\infty}^a d\ell 
\hat P(\ell)
e^{-\hat P(\ell)}\nonumber\\
&=&(1+r)\nu \int_{-\infty}^\infty d\ell\hat Q(\ell)
e^{-e^{\beta(\ell-b)}}e^{-\hat Q(\ell)}\to
(1+r)\nu 
\int_{-\infty}^b d\ell 
\hat Q(\ell)
e^{-\hat Q(\ell)}
\EEA
The male-bachelors energy is
\BEQ
\frac{1}{N}\hat  U_a= a\int_{-\infty}^\infty d\ell\hat\beta
e^{\hat\beta(\ell-c)}e^{-\hat P(\ell)-e^{\hat\beta(\ell-c)}}
\to ae^{-\hat P(a)}
\EEQ
and the female-bachelors energy is
\BEQ
\frac{1}{N}\hat U_b=\nu b\int_{-\infty}^\infty d\ell\hat\beta
e^{\hat\beta(\ell-b)}e^{-\hat Q(\ell)-e^{\hat\beta(\ell-b)}}
\to \nu be^{-\hat Q(b)}
\EEQ
The average fraction of married man is obtained as
\BEA \langle n\rangle &=&\frac{1}{N}\sum_{i=1}^N\langle n_i\rangle 
\nonumber\\
&=&1-\int d\ell
\beta e^{\beta \ell}e^{-e^{\beta(\ell-a)}}e^{-\hat P(\ell)}
\to 1-e^{-\hat P(a)}
\EEA
and likewise the fraction of married woman
\BEA \langle m\rangle &=&\frac{1}{M}\sum_{j=1}^M\langle m_j\rangle
\to 1- e^{-\hat Q(b)}
\EEA
The case of Orland, M\'ezard and Parisi 
is recovered for $\nu=1$ and $a=b=\infty$ 
(infinite energy cost to remain single).

\subsection{A special case at $T=0$}

Let us follow Orland and Mezard-Parisi and take $r=0$. 
We assume equal costs $a=b$
for males and females to remain single. (This restriction is not
necessary; our results can be generalized to the case $a\neq b$.)
For this case eqs. (\ref{PT=0},\ref{QT=0}) become 
\BEQ \label{PQT=0}
Q(\ell)=\int_{-\ell}^a \d y e^{-P(y)}\qquad
P(\ell)=\nu\int_{-\ell}^a \d y e^{-Q(y)}
\EEQ
So that $P=Q=0$ for $\ell\le -a$. One gets
\BEQ P'(\ell)=\nu e^{-Q(-\ell)}\qquad Q'(\ell)= e^{-P(-\ell)}
\EEQ
These equations can be combined in a single equation for $P$
\BEQ
P''(\ell)=P'(\ell)e^{-P(\ell)} \EEQ
which can be integrated once. The solution reads 
\BEQ P(\ell)=
\log\frac{(\gamma-1)e^{\gamma(\ell+a)}+1}{\gamma}
\qquad
Q(\ell)=
\log[\frac{\nu}{\gamma}(\frac{e^{\gamma(\ell-a)}}{\gamma-1}+1)]
\EEQ
where $\gamma$ is a parameter.
It is obvious that $P(-a)=0$, while $Q(-a)=0$ requires that
 $\gamma$ is the largest root of
\BEA
e^{-2\gamma a}&=&(\gamma-1)(\frac{\gamma}{\nu}-1) 
\EEA
From this we derive
\BEQ 
\langle n\rangle =\nu \langle m \rangle =1+\nu-\gamma
\EEQ
In the present case ($r=0$) the energy $U\sim {\cal O}(N^0)$ is 
non-extensive.
The contributions to the energy are at $T=0$:
\BEA
 U_{ab}&=&\sum_{k=1}^\infty \left(\frac{\gamma^{1-k}}{k^2}+
\gamma(1-\frac{\nu}{\gamma})^k[\frac{\ln(\gamma-\nu)}{k}-\frac{1}{k^2}]
\right)\nn
&=&\gamma\ln\frac{\nu}{\gamma}
\ln(\gamma-\nu)
+\gamma\sum_{k=1}^\infty 
\frac{1-(\gamma-\nu)^k}{k^2\gamma^k}
\EEA
\BEQ
\frac{1}{N}\hat U_a=a(\gamma-\nu)\qquad  
\frac{1}{N}\hat U_b=a(\gamma-1)
\EEQ
We can now check some simple cases.
\begin{itemize}
\item 
For $a\to\infty$ and $\nu=1$ one has the Orland-M\'ezard-Parisi case
with $\gamma=1$, so all men and women are coupled. 
If we choose  $\alpha=1/2$, we recover the result
$U_{ab}=\pi^2/12$, while there is no bachelors energy. 

\item
If $a\gg 1$ at fixed $\nu>1$, it follows that $\gamma=\nu$, yielding 
and $ \langle n\rangle =\nu\langle m \rangle =1$ and $ U_a=0$,
saying indeed that all $N$ men are coupled to $N$ out of the $M=\nu N$ women, 
a fraction $1/\nu$ of the total. There is no way to avoid the energy cost
$U_b=a(\nu-1)=(M-N)A$ of the unpairable $M-N$ women, where $A$ is the energy
to pay for each bachelor.
Likewise, for $\nu<1$ there are more men than women, 
and the result $\gamma=1$, 
 $(1/\nu)\langle n\rangle = \langle m \rangle =1$
says that now all women are coupled to a fraction $\nu$ of the men.
This time the surplus of males is responsible for the energy cost
$(N-M)A$. 

\item 
For small $a$ it holds that $U\sim U_{ab}\sim \langle n\rangle\sim a$. 
In the appendix we shall find that the fluctuations scale as $\delta U\sim
a^{3/2}/\sqrt{N}$, implying $\delta U/U\sim \sqrt{a/N}$. As
expected, it vanishes for $a\to 0$.

\item In the limit
 $a\to 0$ it is energetically more advantageous to remain single.
Though pairing may occur at finite $T$ due to entropic reasons, 
the $T=0$ result $\gamma=1+\nu$ indeed implies 
$\langle n\rangle =\langle m \rangle =0$, $\hat U_{ab}=0$,
expressing that no pairs are formed at $T=0$.

\item
Let us finally consider  equally large groups ($\nu=1$) and finite
scaled cost for remaining single, $a=b={\cal O}(1)$.
Then it follows that $\gamma=1+\exp(-\gamma a)$. The result
$<n>=<m>=1-\exp(-\gamma a)$ expresses that a 
finite fraction of bachelors remains, because it is more
advantageous that this fraction of the individuals is not paired.
This aspect remains in the more general case $\nu\neq 1$ and $a\neq b$.

\end{itemize}

\subsubsection{The entropy at low temperatures}

A test for checking the correctness of the replica symmetric approach
is to calculate the leading behavior of the entropy at low
temperatures. For doing that, an expansion in powers of $T$ is needed.

For simplicity we consider the
case $r=0$, $a=b=\infty$, $\nu=1$.
The kernel $B$ from eq. (\ref{Br=}) can be expanded in distribution sense
\BEA \label{BTexp}
B(\ell)&\equiv& 1-J_0(2e^{\b\ell/2})
=\theta(\ell)+a_1\T\delta(\ell)+a_2\T^2\delta'(\ell)+
a_3\T^3\delta''(\ell)\cdots
\EEA
Partial integrations reveal that
\BEQ
a_1=\b\int_{-\infty}^\infty \d\ell \,(B(\ell)-\theta(\ell))=
-2\int_0^\infty\d x\,J_1(x)\log\frac{x}{2}
\EEQ
Similarly
\BEA
a_2=-\b^2\int_{-\infty}^\infty\d\ell\,\ell\,(B(\ell)-\theta(\ell)
-a_1\T\delta(\ell))
   =2\int_0^\infty \d x\, J_1(x)(\log\frac{x}{2})^2 
\EEA 
and 
\BEQ
a_3=
\frac{1}{2}\b^3\int_{-\infty}^\infty\d\ell\,\ell^2\,(B(\ell)-\theta(\ell)
-a_1\T\delta(\ell)-a_2\delta'(\ell))=
-\frac{4}{3}\int_0^\infty \d x\,J_1(x) (\log\frac{x}{2})^3
\EEQ
It holds that ~\cite{Gradstein}
\BEQ \int_0^\infty\d x\left(\frac{x}{2}\right)^\mu J_1(x)=
\frac{\Gamma(1+\frac{1}{2}\mu)}{\Gamma(1-\frac{1}{2}\mu)}
\EEQ
Expansion in powers of $\mu$ yields the needed integrals. One 
obtains
\BEQ a_1=2\gamma_E;\qquad a_2=2\gamma_E^2; \qquad 
a_3=\frac{2}{3}\zeta(3)+\frac{4}{3}\gamma_E^3
\EEQ
where $\gamma_E=0.577215$ is Euler's constant,
and $\zeta(3)=\sum _{k=1}^\infty 1/k^3$ is a Rieman zeta function.

Eq. (\ref{PQell}) reads for the present case
\BEQ P(\ell)=\int_{-\infty}^\infty \d y 
 B(\ell+y)e^{- P(y)}
\EEQ
Expanding
\BEQ
P(y)=P_0(y)+\T P_1(y)+\T^2 P_2(y)+\T^3P_3(y)+
\cdots \EEQ
we find using (\ref{BTexp})
\BEA
P_1(\ell)&=&-\int_{-\ell}^\infty\d y e^{-P_0(y)}P_1(y)+
2\gamma_E \frac{e^\ell} {e^\ell+1}\nn
P_2(\ell)&=&\int_{-\ell}^\infty\d y
e^{-P_0(y)}(-P_2(y)+\frac{1}{2}P_1(y)^2)
-2\gamma_E e^{-P_0(-\ell)}+2\gamma_E^2e^{-P_0(-\ell)}P_0'(-\ell)
\EEA
The solution reads
\BEA
P_0(\ell)&=&\log(e^\ell+1)\\
P_1(\ell)&=&\gamma_E \frac{e^\ell}{e^\ell+1} \\
P_2(\ell)&=&\frac{1}{2}\gamma_E^2 \frac{e^\ell}{(e^\ell+1)^2} 
\EEA
From eq. (\ref{xi-int}) we obtain the entropy
\BEA\label{ST=}
\frac{S}{2N}&=&
\b\int_{-\infty}^\infty \d\ell 
[e^{-e^{\b \ell}}-e^{-P(\ell)}]\nn &=&
\b\int_{-\infty}^\infty \d\ell \left(
 [\theta(-\ell)-e^{-P_0(\ell)}]
+[e^{-e^{\b \ell}}-\theta(-\ell)+\T e^{-P_0(\ell)}P_1(\ell)]
+\T^2[e^{-P_0(\ell)}(P_2(\ell)-\frac{1}{2}P_1(\ell)^2)]\right)
\EEA
The terms calculated so far lead to orders $T^{-1}$, $T^0$ and $T$.
It turns out that all three prefactors vanish. Thus
$S$ is at least of order $T^2$. We have to go one step further.
The equation for $P_3$ reads
\BEA
P_3(\ell)=\int_{-\infty}^\infty\d y e^{-P_0(y)}&\,&\left[
\theta(\ell+y)\{-P_3(y)+P_2(y)P_1(y)-\frac{1}{6}P_1^3(y)\}
+2\gamma_E\delta(\ell+y)\{-P_2(y)+\frac{1}{2}P_1^2(y)\}
\right.
\nn
&\,\,&\left. +2\gamma_E^2\delta'(\ell+y)\{-P_1(y)\}+
(\frac{2}{3}\zeta(3)+\frac{4}{3}\gamma_E^3)\delta''(\ell+y)\right]
\EEA
Some tedious analysis reveals that the solution reads 
\BEQ P_3(\ell)=-\frac{\zeta(3)}{6}\,\frac{e^\ell}{e^\ell+1}
-\frac{\gamma_E^3}{6}\,\frac{e^\ell}{(e^\ell+1)^2}
+(\zeta(3)+\frac{1}{3}\gamma_E^3)\frac{e^\ell}{(e^\ell+1)^3}
\EEQ
The next order contribution to the entropy reads 
\BEQ
\frac{S}{2N}=
\T^2\int_{-\infty}^\infty\d y e^{-P_0(y)}
\{P_3(y)-P_2(y)P_1(y)+\frac{1}{6}P_1^3(y)\}=-\T^2 P_3(\infty)
\EEQ
It follows that
\BEQ S=\frac{1}{3}N\zeta(3)\T^2 \EEQ
The specific heat therefore equals
\BEQ C=\frac{2}{3}N\zeta(3)\T^2 \EEQ
Notice that in $S$ all contributions of order $\gamma_E^3$ have canceled. 
Therefore we could have simplified the calculation by neglecting
$\gamma_E$. This makes it possible to extend the calculations to the
case where $a$ is finite.

\subsection{Random costs for bachelors and $T=0$}

If the costs for bachelors are random with scaled densities
$\rho_1(a)$ and $\rho_2(b)$, then eqs. (\ref{PQT=0}) become
\BEQ Q(\ell)=\int_{-\ell}^\infty \d y R_1(y)e^{-P(y)}\qquad
P(\ell)=\nu\int_{-\ell}^\infty R_2(y) \d y e^{-Q(y)}
\EEQ
where
\BEQ R_1(y)=\int_y^\infty \d a \rho_1(a);
\qquad R_2(y)=\int_y^\infty \d a \rho_2(b)\EEQ
This leads to the differential form
\BEQ Q'(\ell)= R_1(-\ell)e^{-P(-\ell)};\qquad
P'(\ell)=\nu R_2(-\ell)e^{-Q(-\ell)}\EEQ
and
\BEQ
P''(\ell)=-\frac{\rho_2(-\ell)}{R_2(-\ell)}P'(\ell)+R_1(\ell)e^{-P(\ell)}
P'(\ell)
\EEQ
We have not been able to solve these equations for broad
distributions $\rho_1$ and $\rho_2$.
In case that the costs $A_i$ and $B_j$ for remaining 
bachelor take discrete values only, $\rho_1(a)$ and $\rho_2(b)$
are sums of delta-functions. 
Then the solution can be constructed by generalizing our previous
solution in the segments between the $\delta$-functions.

\section{Discussion}

We have considered the bi-partite matching problem, where
members of each group have a wish-list of non-negative real numbers
ranking the possible partners from the other group.
This is a polynomial (P) problem, for which fast algorithms are
available. 
We have  extended previous approaches ~\cite{Orland}\cite{MezardParisi}
to the case where groups have different size, 
members can remain single, and polygamy occurs.
Our results can be extended to the case of random costs for remaining
single or being polygamic, by averaging eq. (\ref{bFN}) over these
variables.  

In section 2 we have given a scaling analysis for the problem.
The interesting regime is where the $\ell$'s, 
the costs for pairings, are low.
We assume that the probability density of the costs of males $\ell^{(a)}$,
scales as a powerlaw, $(\ell^{(a)})^{r_a}$, and the one of females as 
$(\ell^{(b)})^{r_b}$. It is a standard result of statistics
that the probability density for finding
a couple with low cost $\ell=\ell^{(a)}+\ell^{(b)}$
then scales as $\ell^{r}$ with $r=r_a+r_b+1$. The
 exponent relation $r+1=(r_a+1)+(r_b+1)$ expresses
that both partners must have a low cost.

The typical cost can be estimated from $\ell_{typ}^{r+1}=1/N$.
Therefore the energy will scale as $U\sim N\ell_{typ}\sim
N^{r/(r+1)}$. Bachelors and polygamists only play 
an interesting role when their costs 
($A_i, B_j,\mu_{1,i}$ and  $\mu_{2,j}$)
scale as $ U/N\sim N^{-1/(r+1)}$. 
If these costs have a larger magnitude, 
such options do not occur (provided the groups have equal size); 
if they are smaller, no pairs will be formed.
Notice that these scalings of costs for bachelors and polygamists
are not universal to the individual,
but depend on the exponent $r$ entering the distribution
$\ell^r$ of costs for pairings in the whole society.

We have not worked out the role of polygamy at low temperatures.
As it must play a role in case of unequal group sizes ($\nu\neq 1$)
but absence of bachelors ($A=B=\infty$), we expect it generally to
play a non-trivial role down to $T=0$.

In a special case at $T=0$ we generalized previously known 
exact solutions to include for bachelors and unequal group sizes.
It could be checked that if the
typical cost for being bachelor is of the right order of magnitude,
quite a number of members of the groups will remain
bachelor, even if the groups have equal sizes.

It is found that the entropy remains positive
and vanishes quadratically in $T$. This result supports the validity
of the replica symmetric approach.

A calculation of the variance of the free energy is performed
in the Appendix. In principle replica symmetric approaches may
lead to a negative (and thus wrong) prediction for this variance.
The somewhat painful analysis shows that
the replica symmetric Ansatz produces the correct sign.
 In the totally symmetric case of equal group sizes,
equal weighting of males' and females' wishes, and in the
absence of bachelors or polygamy
(with parameters $\nu=1$, $\alpha=1/2$, $A=B=\mu_1=\mu_2=\infty$,
 $r_a=r_b=-1/2$) we derive
${\overline {\delta U^2 }}={0.442878}/{N}$. It has the expected
scaling in $1/N$. The correctness of the sign supports again
the expectation that replica symmetry yields the
correct result for this problem.
In its turn replica symmetry is expected to lead at worst to a
polynomial degeneracy of the ground state.

Our results can be summarized by saying that
whether bachelors occur or not, depends not only on their
intrinsic capacities, or lack thereof, 
but also on global aspects of pair formation in the society.

\acknowledgments
The author's interest for this subject was raised in discussions
with Yi-Cheng Zhang, which is gratefully acknowledged. 
The author  is  grateful for hospitality
at the university of Fribourg, Switserland.
He also thanks the ISI (Torino, Italy) and the ICTP (Trieste, Italy),
where part of this work was done, for hospitality.

\section*{A. Appendix: Variance of the free energy}

The present problem has only been considered within the replica 
symmetric Ansatz. To test the validity of this approach several checks
are possible: a stability analysis, the sign of the 
 entropy or the sign of the second cumulant of the free energy.

The physical free energy should have a positive variance.
This implies that the leading term in $N$ should be non-negative 
(if the leading term vanishes, then the fluctuations have typical 
amplitude less than $\sqrt{N}$).
 It was pointed out by Saakian and Nieuwenhuizen, 
that the ${\cal O}(n^2)$ of $F_n$ can be used to check non-negativity
of the variance~\cite{SaakN}. Indeed, for
$F=-T\log Z$ and $F_n=-T\log {\overline {Z^n}}$,
it holds that
\BEQ\label{dF2=}
\beta^2{\overline{(F-{\overline F})^2\,}}=
-2\beta \lim_{n\to 0}\frac{F_n-n\overline F}{n^2}
\EEQ
In their variational analysis of an interface in a
disordered medium, Saakian and Nieuwenhuizen
 found that the variance is negative if no
replica symmetry breaking is taken into account. For a finite
number of breakings the variance becomes smaller in magnitude,
but remains negative. Only for an infinity of breakings it becomes
zero to leading order.

For simplicity we consider here the symmetric case $A=B$, $\nu=1$,
$\mu_1=\mu_2=\mu$. We first need the $n^2$ term of $z_n^a$, $z_a^{(2)}$,
defined by eq. (\ref{zan=}),
\BEA \label{za2=}
z_a^{(2)}&=&
\sum_{p=1}^\infty \frac{ (-1)^{p}Q_p e^{-\beta\mu p} }{p}(\psi(p)-\psi(1)) 
+\sum_{k=1}^\infty\frac{ e^{-Q(k)}(1-e^{-\beta(A+\mu_1)})^k}
{k}(\psi(k)-\psi(1))
\nn 
&+& \sum_{k=1}^\infty \frac{ e^{-Q(k)}(1-e^{-\beta(A+\mu_1)})^k}{k}
 \sum_{p=1}^\infty\frac{ (-1)^{p-1}Q_p e^{-\beta\mu p} }{p}
-\frac{1}{2}\left(\sum_{k=1}^\infty
\frac{e^{-Q(k)}(1-e^{-\beta(A+\mu_1)})^k}{k}\right)^2
\nn &-& 
\sum_{k=1}^\infty \frac{e^{-Q(k)}(1-e^{-\beta(A+\mu_1)})^k}{k}
\sum_{p=1}^\infty {(-1)^{p-1}Q_pe^{-\beta\mu p} } 
\frac{\Gamma(k+p)}{\Gamma(k)\,p!}(\psi(k+p)-\psi(k))
\EEA
From eqs. (\ref{dF2=}), (\ref{bFn}), (\ref{noverk}), (\ref{ezan=}),
(\ref{zan=}),  and (\ref{QPn})
the variance follows as
\BEA
\frac{\beta^2}{4N} {\overline {\delta F^2 }}
&=&
-\frac{1}{2}\sum_{p=1}^\infty \frac{ (-1)^{p}Q_p^2}{pg_p}(\psi(p)-\psi(1))
+z_a^{(2)}
\EEA
In the first term  we insert the equation of motion (\ref{Qpgen=}). 
We consider the obtained expression in the limit $\mu\to\infty$.
As before, the only possible surviving terms are those where $e^{-\beta\mu}$ 
can  pick up a factor $k$. The relevant terms are
\BEA
\frac{\beta^2}{4N} {\overline {\delta F^2} }&=&
-\frac{1}{2}\sum_{k,p=1}^\infty\frac{(-1)^{p}Q_pe^{-p\beta\mu}\Gamma(k+p)}
{\Gamma(k)p!}(\psi(p)-\psi(1))
\frac{e^{-Q(k)}(1-e^{-\beta(A+\mu_1)})^k}{k}
\nn
&+&\sum_{k=1}^\infty\frac{ e^{-Q(k)}(1-e^{-\beta(A+\mu_1)})^k }
{k}(\psi(k)-\psi(1))
-\frac{1}{2}\left(\sum_{k=1}^\infty 
\frac{e^{-Q(k)}(1-e^{-\beta(A+\mu_1)})^k }{k}\right)^2
\EEA
In the limit $\mu\to\infty$ one obtains an integral representation
involving the variable $\xi=ke^{-\beta\mu}$.
We again add and subtract $e^{-\xi}$ to $e^{-Q(k)}(1-e^{-\beta(A+\mu_1)})^k$
to regularize the small $\xi$-behavior. This leads to
\BEA
\frac{\beta^2}{4N} {\overline {\delta F^2 }}
&=&-\frac{1}{2}\int_0^\infty
\frac{\d\xi}{\xi}e^{-Q(\xi)}e^{-\xi e^{-\beta A}}
 \sum_{p=1}^\infty \frac{ (-1)^{p}Q_p \xi^p} {p!}(\psi(p)-\psi(1)) 
+\int_0^\infty\frac{\d \xi}{\xi}(e^{-Q(\xi)}e^{-\xi e^{-\beta A}}
-e^{-\xi})(\ln\xi+\beta\mu)\nn
&+&(\frac{\beta^2\mu^2}{2}-\gamma_E\beta\mu)+\gamma_E\beta\mu
-\frac{1}{2}\left(\beta\mu+
\int_0^\infty \frac{\d\xi}{\xi}(e^{-Q(\xi)}e^{-\xi e^{-\beta A}}
-e^{-\xi})\right)^2
\EEA
where the last Euler's constant $\gamma_E$ arises from $-\psi(1)$,
while the first comes from an integral representation for 
the sum involving $\psi(k)$.
Notice that all $\mu$-dependent terms cancel, confirming that the
situation without polygamy is well defined. 
Using (\ref{Qeqnxi}) one has finally
\BEA\label{var=}
\frac{\beta^2}{4N} {\overline {\delta F^2 }}
&=&-\frac{1}{2}
\int_0^\infty\frac{\d\xi}{\xi}e^{-Q(\xi)}e^{-\xi e^{-\beta A}} 
\int_0^\infty\frac{\d\eta}{\eta}e^{-Q(\eta)}e^{-\eta e^{-\beta A}}B_2(\xi\eta) 
\nn &+&\int_0^\infty\frac{\d \xi}{\xi}
(e^{-Q(\xi)}e^{-\xi e^{-\beta A}}-e^{-\xi})\ln\xi
-\frac{1}{2}\left(\int_0^\infty \frac{\d\xi}{\xi}
(e^{-Q(\xi)}e^{-\xi e^{-\beta A}}-e^{-\xi})\right)^2
\EEA
where
\BEA\label{B2=}
B_2(\xi\eta)=\sum_{p=1}^\infty \frac{(-1)^{p} pg_p}{p!\,p!}(\xi\eta)^p
(\psi(p)-\psi(1))
\EEA
As can be seen from the first non-vanishing contribution, the 
$p=2$ term, $B_2$ is a positive function.

We now go to $T=0$, so the internal energy equals the free energy.
We calculate $B_2$ for the case $r=0$. We have the scaling
$T\sim \T/N$, $A=a/N$. We  set $\xi=\exp\b\ell$,
$\eta=\exp\b\ell '$ and insert in (\ref{B2=}) the identity 
\BEQ
\psi(p)-\psi(1)= 
-\oint\frac{\d z}{2\pi i z^p}\,\, \frac{\ln(1-z)}{1-z} 
\EEQ
This brings us to a form where the $p$-sum can be carried out, yielding 
\BEA
B_2(\ell+\ell')&=&\T
\oint\frac{\d z}{2\pi i}\,\frac{\ln(1-z)}{1-z}
\left[1-J_0\left(\frac{2e^{\b (\ell+\ell')/2}}{\sqrt{z}}\right)\right]
\EEA
We can deform the $z$-contour and remain with the discontinuity along the
real axis $z\ge 1$,
which yields for $\ell+\ell'\ge 0$
\BEA B_2(\ell+\ell')&=&\T\int_2^\infty \frac{\d z}{z-1}
\left[1-J_0\left(\frac{2e^{\b (\ell+\ell')/2}}{\sqrt{z}}\right)\right]
+{\cal O}(\hat T)\nn
&=&{\ell+\ell'} +{\cal O}(\hat T) \EEA
while $B_2=0$ for $\ell+\ell'\le 0$.

We  calculate the variance of the internal energy for the 
simplest, solvable case $r=0$, $a=b$. 
To leading order, all terms in (\ref{var=})
are proportional to $\b^2\sim \beta^2/N^2$, implying
\BEQ
{\overline {\delta U^2 }}=\frac{2\alpha^{2r_a+2}(1-\alpha)^{2r_b+2}}{N}
(-I_1+2I_2-I_3^2)\EEQ
Since $U\sim N^{r/(r+1)}= N^0$ it follows that $\delta U/U\sim N^{-1/2}$,
as usual. It holds that 
\BEQ
I_1=\int_{-a}^a \d\ell\int_{-\ell}^a
\d\ell'e^{-Q(\ell)-Q(\ell')}(\ell+\ell')
\EEQ
and
\BEQ
I_2=\int_{-a}^0\d\ell\ell(e^{-Q(\ell)}-1)+
\int_0^a\d\ell\ell e^{-Q(\ell)}
\EEQ
and
\BEQ
I_3=\int_{-a}^0\d\ell(e^{-Q(\ell)}-1)+
\int_0^a\d\ell e^{-Q(\ell)}=(\gamma-1)a
\EEQ
For $a\to\infty$ (no bachelors) one has
\BEQ I_1=\int_{-\infty}^\infty\d\ell\ln(1+e^\ell)\,\ln(1+e^{-\ell})
=2*1.202056;
\qquad 
I_2=2\int_0^\infty\ln(1+e^{-\ell})=\frac{\pi^2}{6}=1.644934;
\qquad I_3=0\EEQ
Since $r=0$, boundedness of the entropy at low $T$ implies 
that in this limit
 $I_3=0$, as can also be checked explicitly. Therefore the 
combination $-I_1+2I_2-I_3^2$ is positive. 
In the totally symmetric case (equal group sizes; equal role of males
and females) with $\nu=1$ $\alpha=1/2$, $r_a=r_b=-1/2$ we thus find
\BEQ
{\overline {\delta U^2 }}=\frac{0.442878}{N}\EEQ
(Note that this implies for mono-parted matching problem
with $r=0$, $p(\ell)=\exp(-\ell)$,
considered by M\'ezard and
Parisi\cite{MezardParisi}: ${\overline {\delta U^2 }}=1.771512/N$).
For $a\to 0$ it holds that $\gamma=1+e^{-\gamma a}\sim 2-2a$, and
the prefactor
\BEQ
-I_1+2I_2-I_3^2\approx
-\frac{4a^3}{3}+2(\frac{a^2}{2}-\frac{2}{3}a^3)-(a-2a^2)^2
=\frac{4a^3}{3}+{\cal O}(a^4)\EEQ
is still positive.

It is thus seen that the replica
symmetric prediction for the (free) energy fluctuations
has the correct sign. This supports the expectation that
our replica symmetric solution is exact.

\end{document}